\documentclass[superscriptaddress, amsmath,amssymb, aps, prl, letterpaper, tightenlines, reprint, notitlepages]{revtex4-2}
\usepackage{graphicx}
\usepackage{amsmath}
\usepackage{mathtools}
\usepackage{array}
\usepackage[colorlinks=true,linkcolor=blue,citecolor=blue]{hyperref} 
\usepackage[normalem]{ulem}
\usepackage[dvipsnames]{xcolor}
\usepackage{braket}

\NewDocumentCommand{\evalat}{sO{\big}mm}{%
  \IfBooleanTF{#1}
   {\mleft. #3 \mright|_{#4}}
   {#3#2|_{#4}}%
}

\begin{document}

\title{Autonomous Quantum Error Correction of Spin-Oscillator Hybrid Qubits}
\author{Sungjoo Cho}
\affiliation{NextQuantum Innovation Research Center, Department of Physics and Astronomy, Seoul National University, Seoul 08826, Korea}

\author{Ju-yeon Gyhm}
\affiliation{NextQuantum Innovation Research Center, Department of Physics and Astronomy, Seoul National University, Seoul 08826, Korea}

\author{Hyukjoon Kwon}
\email{hjkwon@kias.re.kr}
\affiliation{School of Computational Sciences, Korea Institute for Advanced Study, Seoul, 02455, South Korea}

\author{Hyunseok Jeong}
\email{jeongh@snu.ac.kr}
\affiliation{NextQuantum Innovation Research Center, Department of Physics and Astronomy, Seoul National University, Seoul 08826, Korea}

\begin{abstract}
We construct an autonomous quantum error-correction protocol for a hybrid qubit that encodes quantum information in spin states and oscillator coherent states. The engineered Lindbladian continuously stabilizes the code space as a steady-state subspace, which results in exponential suppression of the logical phase-error rate as the coherent-state amplitude increases. Remarkably, our protocol employs a single correlated jump operator to suppress phase noise from both the spin and oscillator systems, reducing dissipation-engineering overhead. While exponential phase-noise suppression comes with a linear increase in bit-error rate, we show that concatenating noise-biased hybrid qubits can efficiently suppress both logical errors. Even without concatenation, the strongly biased error profile enables metrological applications that preserve quantum Fisher information beyond the standard quantum limit. Our construction leverages experimentally demonstrated primitives, such as controlled beam-splitter operations readily realizable in trapped-ion systems, offering a practical route toward hardware-efficient quantum error correction.
\end{abstract}
\maketitle

\twocolumngrid
Protecting quantum states from decoherence is a quintessential task in both the fundamental study of quantum information and its practical applications. However, preserving quantum information is more challenging than preserving classical information, as decoherence and measurement itself can easily disturb information encoded in coherence or interference~\cite{RevModPhys.82.1155, RevModPhys.76.1267, nielsen2010quantum}. Quantum error correction (QEC) utilizes syndrome measurements to identify physical errors for correction~\cite{PhysRevA.52.R2493, nielsen2010quantum}, without destroying the quantum information encoded in the code space. Despite remarkable progress on experimental demonstrations of QEC, for example, reporting the break-even between encoded qubits and physical qubits~\cite{paetznick_demonstration_2024} and a sub-threshold scaling~\cite{putterman_hardware-efficient_2025, acharya_quantum_2025, bluvstein_logical_2024}, realizing a large-scale fault-tolerant quantum computing remains a challenging goal. A major technical barrier is that syndrome measurements and decoding require significant resources in both time and physical qubits, which generally introduce additional errors in the system~\cite{haffner_quantum_2008, kjaergaard_superconducting_2020, iolius_decoding_2024}.

Autonomous quantum error correction (AutoQEC)~\cite{doi:10.1098/rspa.1998.0165, PhysRevLett.111.120501} proposes a measurement-free alternative to protect quantum information by designing a passive channel to stabilize the code space. Such a channel can be realized via engineered dissipation~\cite{Cirac_Zoller_PRL, verstraete_quantum_2009, harrington_engineered_2022} by coupling the system to a highly dissipative bath system. AutoQEC protocols have been proposed for both discrete-variable (DV)~\cite{PhysRevA.72.012306, reiter_dissipative_2017, PhysRevA.83.012304, PhysRevLett.128.020403} and continuous-variable (CV)~\cite{mirrahimi_dynamically_2014, lescanne_exponential_2020, PhysRevX.15.011070, xu_autonomous_2023, gottesman_encoding_2001, campagne-ibarcq_quantum_2020, de_neeve_error_2022, Hillman_quantum_2023} quantum systems, where the latter are typically described by harmonic oscillators. In superconducting cavity QED systems,  the generation of noise-biased bosonic cat qubits using AutoQEC, which enables efficient concatenation~\cite{PhysRevX.9.041053, chamberland_building_2022}, was recently realized in experiment~\cite{putterman_hardware-efficient_2025}.

A major challenge in realizing AutoQEC is implementing strong, collective dissipative interactions. Stabilizing DV QEC codes, where the logical information is encoded in the correlations between physical qubits, generally requires multi-qubit interactions. These multi-qubit interactions, along with strong couplings to bath systems to realize engineered dissipation, limit the scalability of AutoQEC for DV systems. For CV QEC codes, instead of multi-qubit interactions, strong nonlinear dissipation in the oscillator, such as two-photon dissipation for stabilizing cat states~\cite{mirrahimi_dynamically_2014}, is necessary~\cite{PhysRevLett.102.120501} for realizing AutoQEC. 

In this Letter, we propose a novel CV–DV hybrid AutoQEC protocol that effectively suppresses phase noise in both DV spin and CV oscillator systems using a single jump operator. We demonstrate that the AutoQEC protocol induces a biased noise profile in the code space spanned by product states of spin states and coherent states, where the logical phase-error rate is exponentially suppressed as the coherent-state amplitude increases. While this comes at the cost of a linearly increasing bit-error rate, we demonstrate that a concatenation of hybrid qubits effectively suppresses both logical bit and phase errors. The noise-biased hybrid states can also be applied to quantum metrology, as the proposed AutoQEC dynamics effectively suppress phase noise, preserving the quantum Fisher information of hybrid entangled states beyond the standard quantum limit.

We highlight that our AutoQEC protocol is experimentally feasible with readily accessible elements in the trapped-ion system, such as controlled beam-splitter interaction~\cite{PhysRevLett.124.170502, jeon2025multimodebosonicstatetomography}, spin-dependent oscillator displacement~\cite{NIST_iontrap}, and a highly dissipative bath. More generally, the CV--DV hybrid construction of AutoQEC can exploit intrinsic DV degrees of freedom, such as the internal atomic levels of trapped-ions~\cite{NIST_iontrap} and the artificial atoms of superconducting circuits~\cite{wallraff_strong_2004}, which can be advantageous compared to the CV-only architecture in terms of performance and implementation. Our hybrid framework coherently combines the infinite redundancy of the CV system with nonlinear operations enabled by coupling to the DV system, creating a distinctive synergy for hardware-efficient error correction.

\textit{Hybrid qubit.---} The hybrid qubit encodes a single logical qubit into a spin-oscillator composite system~\cite{PhysRevA.87.022326, andersen_hybrid_2015}: 
\begin{equation}\label{HES definition}
\begin{aligned}
|+\rangle_\mathrm{L}&:=|+\rangle_s\otimes|+\alpha\rangle_b,\\
    |-\rangle_\mathrm{L}&:=|-\rangle_s\otimes|-\alpha\rangle_b.
\end{aligned}
\end{equation}
Here, $|\pm\rangle_s$ are the $X$-eigenstates of the DV spin system and $|\pm\alpha\rangle_b$ are CV coherent states of the oscillator mode defined by $\hat{D}(\alpha)|0\rangle_b$ with the displacement operator $\hat{D}(\alpha):=e^{\alpha\hat{a}^\dagger-\alpha^*\hat{a}}$. $\hat{a}$ and $\hat{a}^\dagger$ are bosonic annihilation and creation operators, respectively, and without loss of generality, we focus on real-valued $\alpha$. Subscripts $s$ and $b$ denote spin and bosonic oscillator systems, respectively. 

Such a hybrid encoding across CV and DV quantum systems has been explored in photonics~\cite{van_loock_optical_2011, andersen_hybrid_2015, PhysRevA.87.022326, lee_photonic_2026} to overcome the non-orthogonality between small amplitude coherent states ($\langle \alpha\ket{-\alpha}_b \neq 0$) and the complication of their phase operation ($|\pm\alpha\rangle_b\rightarrow\pm|\pm\alpha\rangle_b$), with applications in quantum communication~\cite{bose_quantum_2022, bose_long-distance_2024} and QEC~\cite{PhysRevLett.125.060501, lee_fault-tolerant_2024}. Superposition of these hybrid qubits was also realized in superconducting circuits~\cite{vlastakis_characterizing_2015}, revealing hybrid entanglement between the spin and oscillator systems.

Encoding quantum information from a spin qubit into the hybrid qubit can be implemented by applying the controlled displacement operator $\hat{U}_\mathrm{CD}:=\hat{D}(\alpha\hat{\sigma}_x) = \ket{+}\bra{+}_s \otimes \hat D(\alpha) + \ket{-}\bra{-}_s \otimes \hat D(-\alpha)$ on the initial product state as $    |\psi\rangle_\mathrm{L} = \hat{U}_\mathrm{CD} \left( |\psi\rangle_s\otimes|0\rangle_b \right)$. Decoding of the logical information can be implemented through the inverse unitary $\hat{U}_\mathrm{CD}^\dagger$, enabling measurement on the hybrid qubit through local measurement on the spin subsystem. Alternatively, the encoding and decoding can be realized by controlled parity operation $|+\rangle \langle +|_s \otimes \hat{\mathbb{I}}_b +|-\rangle{}\langle-|_{s} \otimes \hat\Pi_b$ acting on $|\psi\rangle_s\otimes|\alpha\rangle_b$, where $\hat\Pi_b = (-1)^{\hat{n}}$.

The logical $X$ operation ($\hat X_L$) is implemented by applying a Pauli $X$ operation $\hat{\sigma}_x$ on the spin qubit. The logical $Z$ operation ($\hat Z_L$) can be implemented by either a product of the spin phase-flip and the oscillator parity operator, i.e., $\hat{\sigma}_z\otimes \hat\Pi_b$, or a spin-dependent displacement followed by a phase-flip, i.e., $\hat \sigma_z \hat D(-2\alpha \hat\sigma_x)$. The displacement-type logical $Z$ operation can be realized by using a spin-dependent force~\cite{PhysRevLett.94.153602, johnson_ultrafast_2017} that is native to the trapped-ion systems and also realizable for superconducting systems ~\cite{PhysRevLett.122.080502}. For entangling operations between multiple logical qubits, an $XX$ interaction between the DV subsystems $e^{-i\frac{\theta}{2}\hat{\sigma}_x\otimes\hat{\sigma}_x}$ is sufficient to form a universal gate set when combined with logical single-qubit gates.

\begin{figure}[b]
\includegraphics[width=\linewidth]{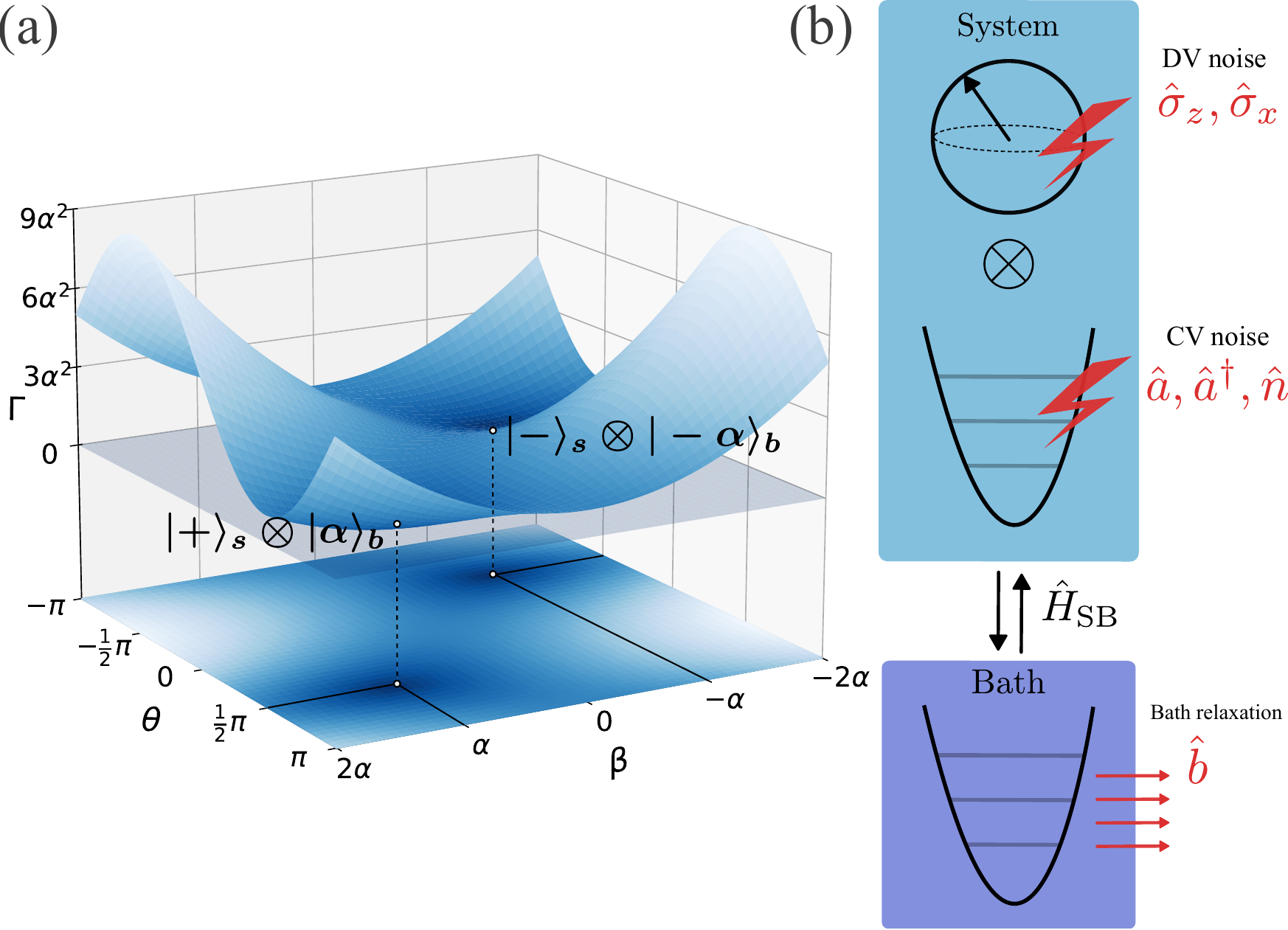}
\caption{(a) $\Gamma(\psi)$ of ansatz states $|\theta\rangle_s\otimes|\beta\rangle_b$ as a function of $\beta$ and $\theta$. The only pure stationary states having $\Gamma=0$ are $|\psi\rangle_\mathrm{L}\in\mathrm{span}\{|\pm\rangle_\mathrm{L}\}$. (b) Schematic diagram of the AutoQEC dynamics of a hybrid qubit.}
\label{Hybrid_schematics} 
\end{figure}
\textit{AutoQEC of the hybrid qubit.---} We consider the dynamics of the spin-oscillator hybrid system described by the Lindblad master equation~\cite{lindblad1976generators}:
$$
    \frac{d{\hat{\rho}}}{dt} = \mathcal{L}(\hat{\rho} )
    := -i[\hat{H}, \hat{\rho}] 
    + \sum_{k}\kappa_k\mathcal{D}[\hat{L}_k](\hat{\rho}),
$$
where $\hat{H}$ denotes the Hamiltonian of the system, $\mathcal{D}[\hat{L}](\hat{\rho}) 
:= \hat{L}\hat{\rho}\hat{L}^\dagger 
- \frac{1}{2}\{ \hat{L}^\dagger \hat{L},\  \hat{\rho}\}$ describes dissipation given by a jump operator $\hat L$, and $\kappa_k$ are the dissipation rates for each $\hat L_k$. Also, $[\cdot,\cdot]\ (\{\cdot, \cdot\})$ denotes the canonical (anti-)commutator. We decompose the quantum Liouvillian into two parts,
$$
    {\cal L} = {\cal L_{\mathrm{E}}} + {\cal L}_\mathrm{R},
$$
where ${\cal L_{\mathrm{E}}}$ describes the uncontrollable noise of the system that induces errors, and ${\cal L}_\mathrm{R}$ is engineered dissipation that stabilizes the code space and recovers it from errors. To design AutoQEC for the logical qubit spanned by the hybrid states described in Eq.~\eqref{HES definition}, we construct $\mathcal{L}_\mathrm{R} := \kappa_\mathrm{R}\mathcal{D}[\hat{R}]$ with the correlated jump operator,
\begin{equation}\label{recovery}
\begin{aligned}
    \hat{R} &:= \hat{\sigma}_z(\alpha-\hat{\sigma}_x\otimes\hat{a}).
\end{aligned}
\end{equation}
We note that for such recovery Liouvillian $\mathcal{L}_R$, every density operator $\hat{\rho}_\mathrm{L}$ in the $2$-dimensional code space spanned by $\{|\pm\rangle_\mathrm{L}\}$) becomes a stationary point, i.e., $\mathcal{L}_R (\hat{\rho}_\mathrm{L})=0$.

The code space can be interpreted as the ground state manifold of the Hamiltonian $\propto \hat{R}^\dagger\hat{R}$, where the stability of states $\ket\psi$ outside the code space can be characterized by the effective potential $\Gamma(\psi) = \bra \psi \hat{R}^\dagger\hat{R} \ket\psi$. For a coherent state ansatz $|\psi\rangle:= |\theta\rangle_s\otimes| \beta\rangle_b$, with a spin state $|\theta\rangle$ such that $\langle\hat{\sigma}_x\rangle_\theta=\sin{\theta}$, we obtain
$$
    \Gamma(\psi) = \alpha^2-2\alpha {\rm Re}[\beta]\sin{\theta}+|\beta|^2,
$$
which is minimized for ${\rm Im}[\beta] = 0$ and forms an attractive potential around $(\theta,\beta) = (\pm \pi/2, \pm\alpha)$ (see Fig.~\ref{Hybrid_schematics}(a)).

As having a $2$-dimensional stationary code space, the recovery dynamics has four conserved quantities, $\hat J_{0,1,2,3}$ such that ${\cal L}_\mathrm{R}^\dagger (\hat J_{0,1,2,3}) = 0$~\cite{Sym_consv_VValbert}, including the trivial one $\hat J_0 = \mathbb{\hat I}_s \otimes \mathbb{\hat I}_b$ from the trace-preserving condition. Also, a strong symmetry under the hybrid parity $\hat{\Pi}_h:=\hat{\sigma}_z\otimes \hat\Pi_b$, satisfying $[\hat{\Pi}_h,\hat{R}]=[\hat{\Pi}_h,\hat{R}^\dagger]=0$, leads to $\hat J_3 = \hat{\Pi}_h$. The remaining two conserved quantities are $\hat J_1 = \ket{+}\bra{+}_s \otimes \hat K - \ket{-}\bra{-}_s \otimes \hat \Pi_b \hat K \hat \Pi_b$ and $\hat J_2 = \hat J_3 \hat J_1 = -\ket{+}\bra{-}_s \otimes \hat K \hat \Pi_b + \ket{-}\bra{+}_s \otimes \hat \Pi_b \hat K$, where $\hat K$ is the solution of 
$$
    (\hat a^\dagger - \alpha) \hat{\Pi}_b \hat{K} \hat{\Pi}_b (\hat a - \alpha) + \frac{1}{2} \left\{ (\hat a^\dagger - \alpha)(\hat a - \alpha), \hat{K}\right\}  = 0,
$$
with normalization $\bra \alpha \hat K \ket \alpha_b = 1$ (see Supplemental Material (SM)~\cite{supple} for details). We note that ${\rm Tr} \left[ \hat J_{1,2,3}^\dagger \hat\rho_{\rm L}\right]$ for $\hat \rho_{\rm L}$ in the code space corresponds to its logical Pauli expectation values in $x$, $y$, and $z$-axes, respectively.

\begin{table}[t]
    \centering
    \begin{tabular}{c | c | c}
        \hline
         Physical noise & Bit error rate ($\gamma_X)$ & Phase error rate ($\gamma_Z$)\\
         \hline 
            $\kappa_{\hat{\sigma}_x} \mathcal{D}[\hat{\sigma}_x]$ & $\kappa_{\hat{\sigma}_x}$ & 0\\
            $\kappa_{\hat{\sigma}_z} \mathcal{D}[\hat{\sigma}_z]$ & 0 & $\frac{\kappa_{\hat{\sigma}_z}}{1+2I(\alpha)}\sim\frac{\kappa_{\hat \sigma_z}}{\sqrt{2\pi}\alpha}e^{-2\alpha^2}$\\
            $\kappa_{\hat{a}} \mathcal{D}[\hat{a}]$ & $\kappa_{\hat{a}}\alpha^2$ & 0\\
            $\kappa_{\hat{a}^\dagger} \mathcal{D}[\hat{a}^\dagger]$ & $\kappa_{\hat{a}^\dagger}(\alpha^2+1)$ & $\frac{\kappa_{\hat{a}^\dagger}}{1+2I(\alpha)}\sim\frac{\kappa_{\hat{a}^\dagger{}}}{\sqrt{2\pi}\alpha}e^{-2\alpha^2}$\\
            $\kappa_{\hat{n}}\mathcal{D}[\hat{n}]$ & 0 & $\frac{\kappa_{\hat{n}}\alpha^2}{1+2I(\alpha)}\sim\frac{\kappa_{\hat n}}{\sqrt{2\pi}}\alpha e^{-2\alpha^2}$\\
        \hline
    \end{tabular}
    \caption{First order approximation of logical bit (phase) error rates $\gamma_\mathrm{X}\ (\gamma_\mathrm{Z})$ induced by each physical error channel in the rapid dissipation limit ($\kappa_\mathrm{R}/\kappa_\mathrm{E}\gg1$). Here, we define $I(\alpha) := \alpha \sqrt{\pi/2} e^{2\alpha^2} \mathrm{erf} (\sqrt{2}\alpha)$ with $\mathrm{erf}(x):=2/\sqrt{{\pi}}\int_{0}^{x}dte^{-t^2}$.}
    \label{tab:error_rate_summary}
\end{table}

\textit{Dynamics of the hybrid qubit under noise.---}
We analyze the logical error rates under local Markovian errors given by the following Liouvillian:
\begin{equation} \label{eq:noise_lioubillian}
    \mathcal{L}_{\mathrm{E}}
    := \mathcal{L}_\mathrm{th} + \kappa_{\hat{n}}\mathcal{D}[\hat{n}]+\kappa_{\hat{\sigma}_x}\mathcal{D}[\hat{\sigma}_x]+\kappa_{\hat{\sigma}_z}\mathcal{D}[\hat{\sigma}_z].
\end{equation}
Here, $\mathcal{L}_\mathrm{th}:= \kappa_{\hat a} \mathcal{D}[ \hat a] + \kappa_{\hat a^\dagger} \mathcal{D}[ \hat a^\dagger]$ with $\kappa_{\hat a} = \kappa_\mathrm{th}(n_\mathrm{th}+1)$ and $\kappa_{\hat a^\dagger} = \kappa_\mathrm{th}n_\mathrm{th}$ describes the thermal noise channel for a bosonic system coupled to a thermal environment state with the mean excitation number $n_\mathrm{th}$. A classical field noise model corresponding to the infinite-temperature bath weakly coupled to the system, i.e., $\mathcal{L}_\mathrm{th}=\gamma_\mathrm{th}(\mathcal{D}[\hat{a}]+\mathcal{D}[\hat{a}^\dagger{}])$ with $\kappa_\mathrm{th}n_\mathrm{th}=\gamma_\mathrm{th}$ in the limit of $n_\mathrm{th}\rightarrow \infty$, is widely accepted for trapped-ion system~\cite{RevModPhys.87.1419}. For a circuit-QED system \cite{putterman_hardware-efficient_2025}, the cryogenic environment of the superconducting circuit corresponds to a mean thermal photon number $n_\mathrm{th}\approx10^{-3}$, indicating a loss-dominant thermal noise channel.

We illustrate how a single jump operator $\hat R$ in Eq.~\eqref{recovery} can suppress phase noise from both qubit and oscillator systems, by analyzing the effect of the recovery operation on the logical state $\ket\psi_\mathrm{L}$ under jump operators corresponding to the spin and oscillator systems' phase noise. For the spin phase flip $\hat \sigma_z \otimes \mathbb{\hat I}_b$, we have 
$$
    (\hat \sigma_z \otimes \hat{\mathbb{I}}_b )\ket \psi_\mathrm{L} \xrightarrow[{\rm recovery}]{\hat R}  \hat R(\hat \sigma_z \otimes \hat{\mathbb{I}}_b ) \ket\psi_\mathrm{L}= 2 \alpha \ket\psi_\mathrm{L},
$$
so that the logical state is recovered to the code space. For the oscillator noise $\mathbb{\hat I}_s \otimes \hat{n}$, we observe that
$$
\begin{aligned}
    (\mathbb{\hat I}_s \otimes \hat{n}) \ket \psi_\mathrm{L}
    &\xrightarrow[{\rm recovery~1}]{\hat R} \hat R (\mathbb{\hat I}_s \otimes \hat{n}) \ket \psi_\mathrm{L} = -\alpha (\hat \sigma_z \otimes \hat{\mathbb{I}}_b ) \ket\psi_\mathrm{L} \\
    & \xrightarrow[{\rm recovery~2}]{\hat R}  -\alpha \hat R (\hat \sigma_z \otimes \hat{\mathbb{I}}_b ) \ket\psi_\mathrm{L} = -2 \alpha^2 \ket\psi_\mathrm{L}, 
\end{aligned}
$$
where the oscillator's phase noise is converted into the spin phase noise and then recovered to the code space. Such a noise suppression mechanism also applies to induced phase errors from thermal noise, where a more detailed structure of the hybrid AutoQEC dynamics under noise can be found in End Matter. 

More rigorously, using first-order perturbation theory in the limit $\kappa_{\rm R} \gg \kappa_{\hat \sigma_z}, \kappa_{\hat n}$ and the conserved quantity $\hat J_{1}$, we demonstrate that the logical phase-error rate $\gamma_Z$ is exponentially suppressed as $\alpha^2$ increases (see Table~\ref{tab:error_rate_summary} and SM~\cite{supple} for details). Here, $\gamma_Z$ is defined as the population decay rate of the eigenstate initially prepared in $\ket{+}_{\rm L}$, i.e., $p_{Z}(t) = {\rm Tr} \left[ \ket{+}\bra{+}_{\rm L} e^{(\mathcal{L}_\mathrm{R}+\mathcal{L}_\mathrm{E}) t} (\ket{+}\bra{+}_{\rm L}) \right] = (1+e^{-2\gamma_{Z} t})/2$. Especially for the bosonic phase noise, our protocol provides consistently better noise suppression $(\gamma_Z \approx \alpha e^{-2\alpha^2})$ compared to the cat code with two-photon substraction ($\gamma_Z \approx \alpha^2 e^{-2\alpha^2}$)~\cite{xu_autonomous_2023} with factor $\alpha$.

Meanwhile, the recovery dynamics is not able to recognize the qubit bit-flip noise $\hat \sigma_x$, as $[\mathcal{D}[\hat{\sigma}_x], \mathcal{D}[\hat{R}]]=0$. The thermal loss $\hat{a}, \hat{a}^\dagger$ translates into a logical bit-flip rate that scales linearly with $\alpha^2$~\cite{supple}, since loss acts as a logical $X$-operation, i.e., $(\hat{\mathbb{I}}_s \otimes \hat{a}) \ket{\psi}_\mathrm{L} = \alpha\hat{X}_\mathrm{L} |\psi\rangle_\mathrm{L}$, similarly to the AutoQEC applied to the cat code~\cite{mirrahimi_dynamically_2014, lescanne_exponential_2020}. We summarize the logical $X$ and $Z$ error rates for each element in Eq.~\eqref{eq:noise_lioubillian} in Table~\ref{tab:error_rate_summary}.

\begin{figure}[t]
    \centering
    \includegraphics[width=\linewidth]{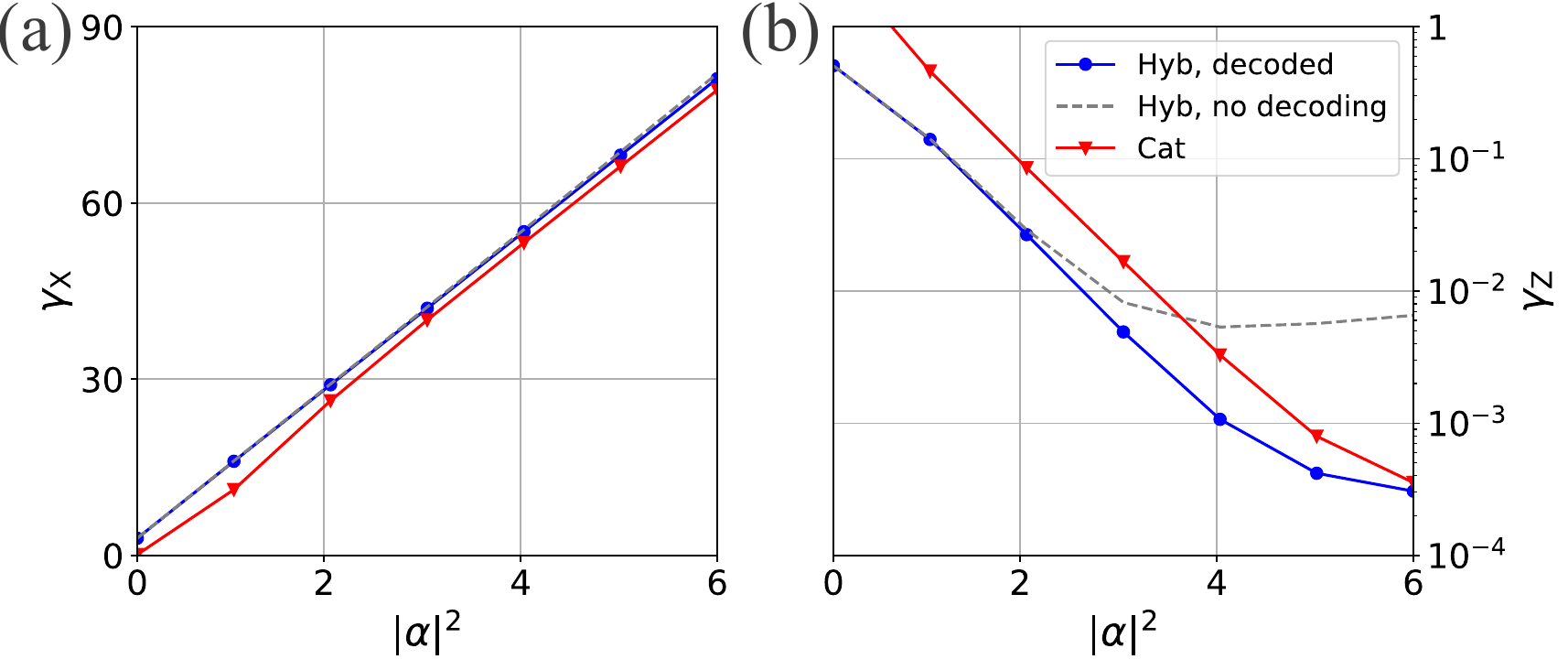}
    \caption{Logical (a) bit error rates and (b) phase error rates of the hybrid qubit and the cat qubit under AutoQEC dynamics. The simulation parameters can be found in End Matter.}
    \label{lifetime_tradeoff}
\end{figure}

Figure~\ref{lifetime_tradeoff} depicts a numerical simulation of logical error rates using realistic experimental parameters given in End Matter, which clearly demonstrates an exponential–linear tradeoff between the phase–bit error rates. The constant gap in the logical bit error rate between the hybrid qubit and cat qubits described in Fig.~\ref{lifetime_tradeoff}(a) stems from the spin noise $\mathcal{D}[\hat{\sigma}_x]$, which only applies to the hybrid qubit. We also note that the residual phase errors can be further reduced by utilizing the decoding operation $\hat{U}_\mathrm{CD}^\dagger$. Since the bosonic mode after disentangling is irrelevant, any superposition of $\hat{U}_\mathrm{CD} (|\psi\rangle_s \otimes |n\rangle_b )$ can contribute to the measurement result without applying corrections to their bosonic modes (equivalent to the gauge group arising in degenerate stabilizer codes~\cite{PhysRevLett.95.230504}).
Our numerical simulation shows that our AutoQEC protocol outperforms the AutoQEC of the cat code ($|\mu\rangle_\mathrm{cat} := |\alpha\rangle_b+(-1)^{\mu}|-\alpha\rangle_b,~ \mu\in\{0,1\}$) using the two-photon dissipation~\cite{mirrahimi_dynamically_2014} for any coherent state amplitudes with $\alpha^2 \leq 6$ with the same noise parameters.

\begin{figure}[t]
    \centering
    \includegraphics[width=.9\linewidth]{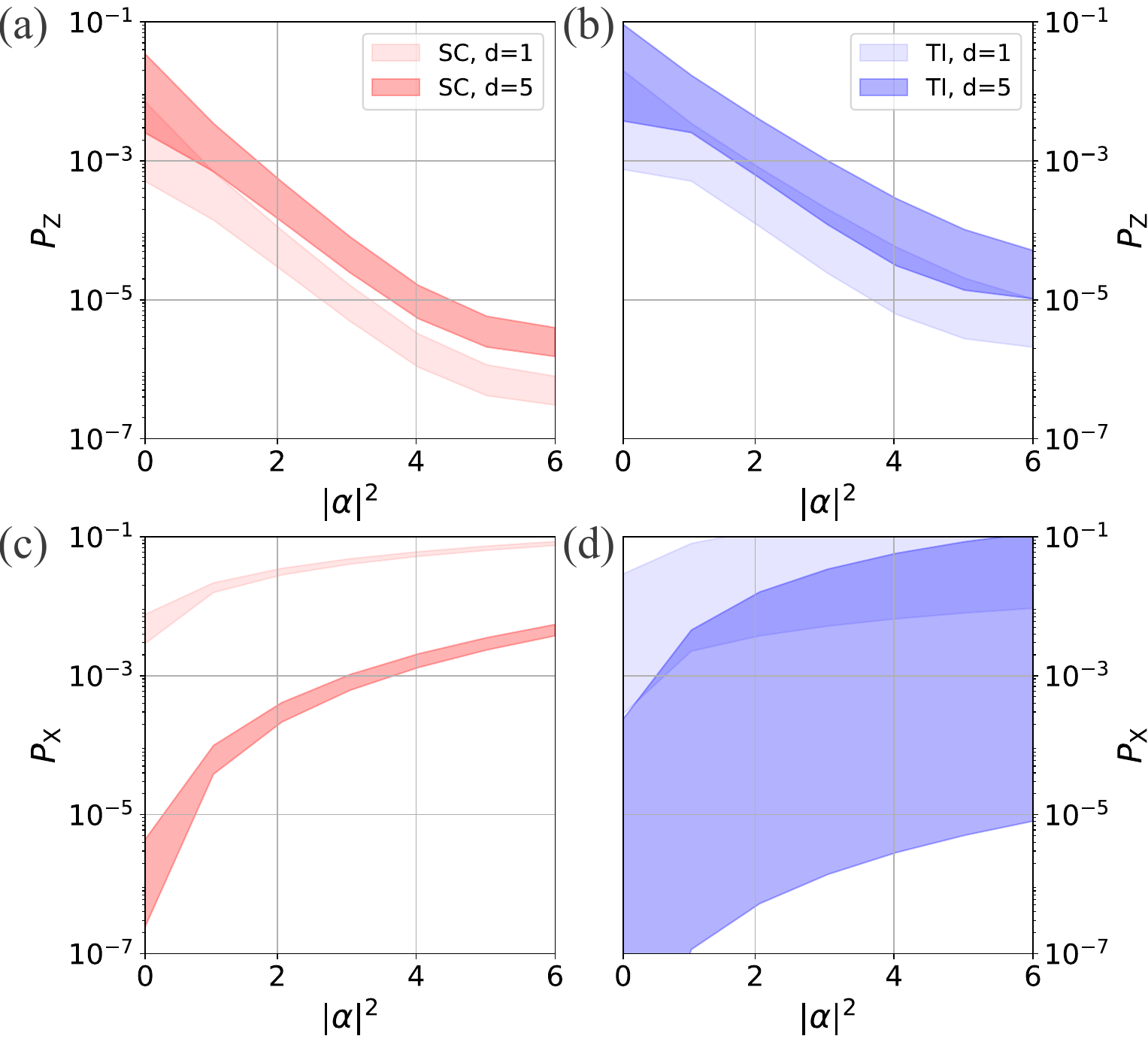}
    \caption{The error probabilities of concatenated hybrid qubits ($d=5$) after $t =1\ \mathrm{ms}$ for trapped-ions~\cite{PhysRevLett.117.060504} and $t =1\ \mathrm{\mu s}$ for superconducting circuits~\cite{PhysRevLett.115.137002}. Noise parameters for each platform is given in End Matter.}
    \label{fig_rep}
\end{figure}

\textit{Concatenation of the hybrid qubit.---} The bit errors amplified from the proposed AutoQEC can be further suppressed by the concatenation of the hybrid qubit using a conventional QEC code. Such a proposal has been made for cat codes~\cite{PhysRevX.9.041053} and was recently realized in superconducting systems~\cite{putterman_hardware-efficient_2025}. As the simplest example, we consider a repetition code, $|\mu\rangle_\mathrm{rep} := |\mu\rangle_\mathrm{L}^{\otimes d},~\mu\in\{0,1\}$ with distance $d$. The bit repetition code suppresses the bit errors, but makes the logical information more susceptible to phase-flip errors as each phase-flip in $\ket\mu_\mathrm{L}$ induces a logical phase error. More precisely, the logical error probabilities ($P_\mathrm{X}, P_\mathrm{Z}$) in terms of unconcatenated error probabilities $(q_\mathrm{X}, q_\mathrm{Z})$ for each $\ket\mu_\mathrm{L}$ is given as,
$$
\begin{aligned}
    &P_\mathrm{Z} = \frac{1}{2}(1-(1-2q_\mathrm{Z})^d),\\
    &P_\mathrm{X} = \sum_{k=\lfloor \frac{d+1}{2}\rfloor}^{d} {d \choose k} q_\mathrm{X}^k(1-q_\mathrm{X})^{d-k}.
\end{aligned}
$$
Especially for $q_\mathrm{X},q_\mathrm{Z} \ll 1$, $P_\mathrm{Z}\propto dq_\mathrm{Z},\ P_\mathrm{X}\propto q_\mathrm{X}^{(d+1)/2}$. This compensates the exponential-linear tradeoff between the phase-bit error rates in hybrid qubits $\ket\mu_{\mathrm{L}}$ via the proposed AutoQEC dynamics, which leads to efficient suppression of both logical phase and bit errors after the concatenation (see Fig.~\ref{fig_rep}) 

\textit{AutoQEC for noise-suppressed quantum metrology.---} Exponential phase-noise suppression of the hybrid qubit via the proposed AutoQEC dynamics can also be applied to quantum metrology. We particularly focus on the estimation of the displacement~\cite{gilmore_quantum-enhanced_2021} with parameter $\beta$ for $\hat D(i \beta)$ acting on the oscillator system in the $p$-direction. The quantum Cram\'er-Rao bound (QCRB)~\cite{helstrom_minimum_1967} determines the ultimate sensitivity of a given probe state $\hat \rho$ in terms of quantum Fisher information ${\cal F}_Q(\hat\rho)$ as, $(\Delta \beta)^2 \geq 1/{\cal F}_Q(\hat\rho)$ (see End Matter for details). 

For the noiseless case, a spin-oscillator entangled state in the code space, $\ket{0}_\mathrm{L} = \frac{1}{\sqrt{2}} \left(\ket{+}_s \otimes \ket{\alpha}_b + \ket{-}_s\otimes\ket{-\alpha}_b\right)$ achieves the precision $(\Delta \beta)^2 \geq 1/{{\cal F}_Q(\ket{0}\bra{0}_\mathrm{L})} = 1/(16\alpha^2+4)$, surpassing the standard quantum limit (SQL), $(\Delta\beta)^2 \geq 1/4$, whenever $\alpha^2 > 0$.
\begin{figure}[t]
    \centering
    \includegraphics[width=\linewidth]{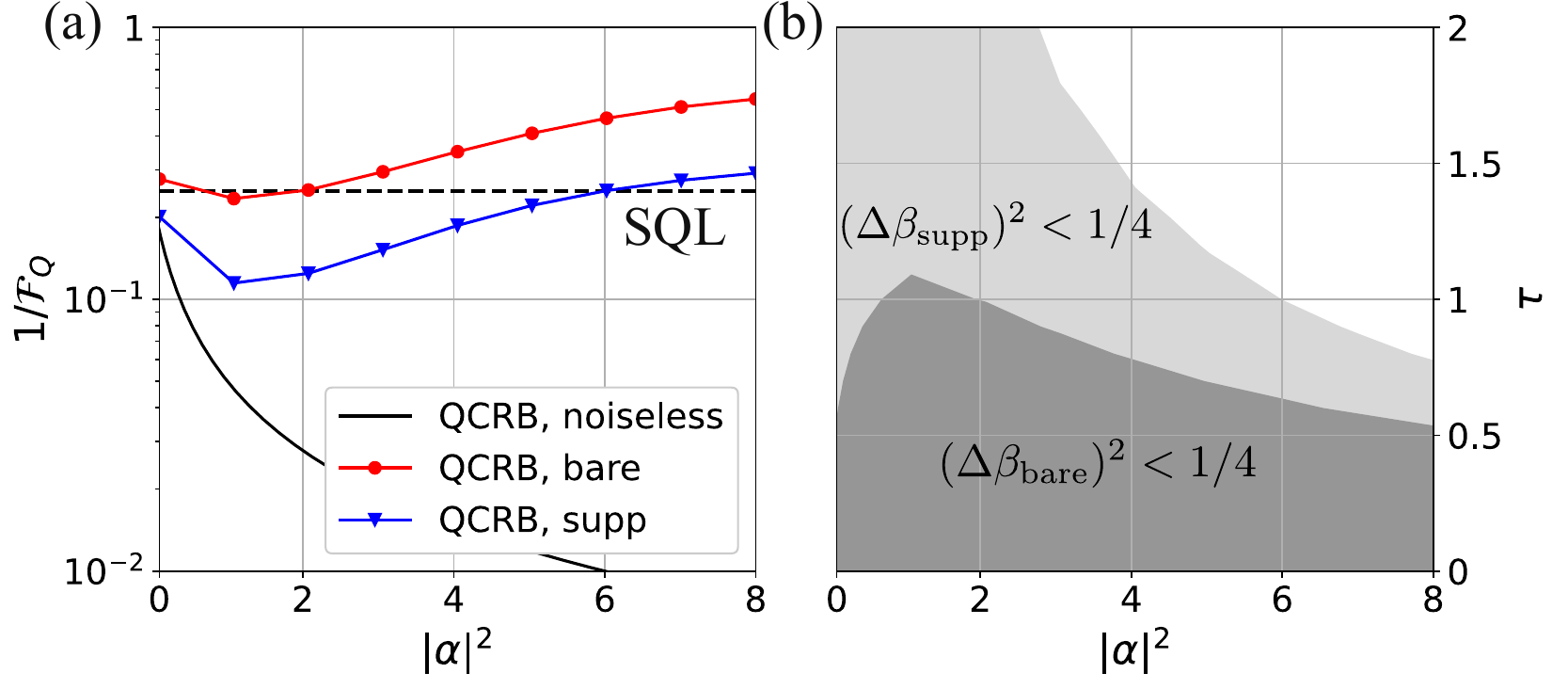}
    \caption{(a) QCRB of the ideal probe state $\ket{0}\bra{0}_\mathrm{L}$ and the noise-suppressed ($\hat\rho_{\rm supp}$) and unsuppressed ($\hat \rho_{\rm bare}$) states exposed to noise for $\tau= 1\ \mathrm{ms}$. The noise-suppressed probe state (blue triangle) consistently outperforms the bare probe state (red circle). (b) Regime where sub-SQL displacement sensitivity is achievable using $\hat \rho_{\rm bare}$ (dark grey) and $\hat\rho_{\rm supp}$ (light grey). Experimental parameters are considered for ion-trap systems (see End Matter).}
    \label{fig:displacement_sensing}
\end{figure}
While such an advantage can be easily washed out when the probe state is exposed to the noise, our AutoQEC dynamics effectively protects the probe state against external noise to maintain sub-SQL~\cite{kwon2025restoring}. Figure~\ref{fig:displacement_sensing} shows that, for a given time $\tau = 1~{\rm ms}$, the QFI of the noise-suppressed probe state, $\hat{\rho}_{\rm supp} := e^{({\cal L}_\mathrm{E} + {\cal L}_\mathrm{R}) \tau}(\ket{0}\bra{0}_\mathrm{L})$, is more robust against noise than that of the unsuppressed state, $\hat{\rho}_{\rm bare} := e^{{\cal L}_\mathrm{E} \tau}(\ket{0}\bra{0}_\mathrm{L})$.

\textit{Experimental implementation.---}
The recovery Liouvillian $\mathcal{L}_\mathrm{R}$ can be realized by interacting the spin-oscillator system with an additional dissipative mode~\cite{Cirac_Zoller_PRL, verstraete_quantum_2009} (see also Fig. \ref{Hybrid_schematics}(b)).
The following Hamiltonian implements the dissipation $\kappa_\mathrm{R} \mathcal{D}[\hat{R}]$ for stabilizing the hybrid qubit, 
$$
    \hat{H}_\mathrm{SB} = g\hat{R}\hat{b}^\dagger+\mathrm{h.c.} = g[ \alpha\hat{\sigma}_z\hat{b}^\dagger-i\hat{\sigma}_y\hat{a}\hat{b}^\dagger
]+\mathrm{h.c.},
$$
where $\hat{b}$ is the annihilation operator of the bath mode under strong dissipation $\gamma_b {\cal D}[\hat b]$ and $\mathrm{h.c.}$ denotes the Hermitian conjugate. In the strong dissipation limit $(\gamma_b \gg g)$, adiabatic elimination translates the system-bath coupling into an effective dissipation rate $\kappa_\mathrm{R} \approx 4g^2/\gamma_b$~\cite{reiter_effective_2012}. $\hat{H}_\mathrm{SB}$ consists of a spin-dependent displacement~\cite{PhysRevLett.94.153602, johnson_ultrafast_2017, PhysRevLett.122.080502} and a controlled beam-splitter interaction, which has been realized in trapped-ion systems for spin $z$-direction~\cite{PhysRevLett.124.170502} and an arbitrary direction of the $xy$-plane~\cite{jeon2025multimodebosonicstatetomography}. In the superconducting system, the controlled beam-splitter term in the $\hat{H}_\mathrm{SB}$ can be, in principle, realized by utilizing the nondegenerate three-wave mixing from native nonlinear elements of circuit QED~\cite {PhysRevX.15.011070, frattini_three-wave_2021}. Strong, continuous relaxation of the motional mode can be realized in trapped-ion systems through Raman sideband cooling~\cite{PhysRevResearch.5.023022} or cooling by electromagnetically induced transparency~\cite{PhysRevLett.125.053001}. In a superconducting system, a lossy bath mode can be realized using superconducting quantum interference devices~\cite{PhysRevX.15.011070, frattini_three-wave_2021}.

\textit{Remarks.---} We have proposed an AutoQEC architecture for stabilizing a hybrid qubit encoded in spin states and oscillator coherent states. Our formalism provides measurement-free exponential suppression of phase noise from both the spin and oscillator systems via a single correlated jump operator, which is implementable with elements already realized in trapped-ion systems. The resulting noise-biased hybrid qubits provide a promising resource for concatenated quantum error correction and quantum metrology.

Our hybrid AutoQEC framework offers a hardware-efficient pathway for harnessing the advantages of the infinite-dimensional structure of bosonic modes activated by spin-mediated controllability, enabling capabilities inaccessible to approaches based on DV or CV systems alone. This new framework also opens a wide range of future research directions, including designing more efficient noise-biased QEC codes for concatenation~\cite{hann_hybrid_2025}, developing error-bias-preserving gate operations~\cite{puri_bias-preserving_2020, xu_engineering_2022} for hybrid qubits, and engineering CV--DV interfaces for efficient physical implementation.

\begin{acknowledgements}
This work was supported by the National Research Foundation of Korea (NRF) grants funded by the Korea government (MSIT) (No.~RS-2024-00413957 and No. RS-2024-00438415), and the Institute of Information \& Communications Technology Planning \& Evaluation (IITP) grant funded by the Korea government (MSIT) (IITP-2025-RS-2020-II201606 and  IITP-2025-RS-2024-00437191). H.K. is supported by the KIAS Individual Grant No. CG085302 at Korea Institute for Advanced Study.

\end{acknowledgements}

\bibliographystyle{no_article_note}
\bibliography{bibtextest}

\newpage
\onecolumngrid
\section{End Matter}
\twocolumngrid

\noindent \textbf{AutoQEC dynamics under noise.} We define a couple of bases that span the spin-oscillator system using the controlled-displacement unitary operator $\hat{U}_\mathrm{CD}= \hat{D}(\alpha\hat{\sigma}_x)$, as follows:
$$
\begin{aligned}
    |\pm, \overrightarrow{n}\rangle&:=\hat{U}_\mathrm{CD}|\pm\rangle_s\otimes|n\rangle_b = \hat{D}(\pm\alpha)|\pm\rangle_s\otimes|n\rangle_b,\\
    |\pm, \overleftarrow{n}\rangle&:=\hat{U}^\dagger_\mathrm{CD}|\pm\rangle_s\otimes|n\rangle_b = \hat{D}(\mp\alpha)|\pm\rangle_s\otimes|n\rangle_b.
\end{aligned}
$$
We note that the both bases satisfy the orthogonality relation, $\langle\mu,\overrightarrow{n}|\nu,\overrightarrow{m}\rangle=\delta_{\mu\nu}\delta_{nm},\  \langle\mu,\overleftarrow{n}|\nu,\overleftarrow{m}\rangle=\delta_{\mu\nu}\delta_{nm}$ and the completeness relation, $\sum_{\mu,n}|\mu,\overleftarrow{n}\rangle\langle\mu,\overleftarrow{n}|=\sum_{\mu,n}|\mu,\overrightarrow{n}\rangle\langle\mu,\overrightarrow{n}|=\hat{\mathbb{I}}_s\otimes\hat{\mathbb{I}}_b$.

\begin{figure}[t]
    \centering
    \includegraphics[width=0.5\linewidth]{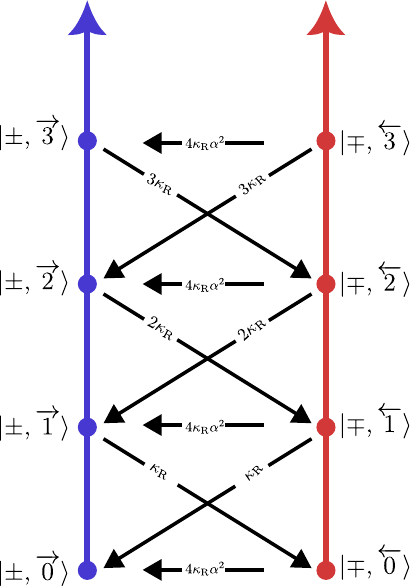}
    \caption{Approximate Markov diagram of states up to excitation. The coherence of post-jump states is neglected. Numbers in arrows correspond to the rates of the corresponding jumps. Every transition involves a phase flip.}
    \label{fig:Markov_diagram}
\end{figure}

We calculate the action of jump operators in the set ${\cal E}$ on relevant pure states. For the spin system's jump operators, we have 
$$
\begin{aligned}
    \hat{\sigma}_x|\pm,\overrightarrow{n}\rangle&=\pm|\pm,\overrightarrow{n}\rangle,\\
    \hat{\sigma}_x|\pm,\overleftarrow{n}\rangle&=\pm|\pm,\overleftarrow{n}\rangle,\\
    \hat{\sigma}_z|\pm,\overrightarrow{n}\rangle&=|\mp,\overleftarrow{n}\rangle,\\
    \hat{\sigma}_z|\pm,\overleftarrow{n}\rangle&=|\mp,\overrightarrow{n}\rangle.
\end{aligned}
$$
For the oscillator system's jump operator, we have 
$$
\begin{aligned}
    \hat{a}|\pm,\overrightarrow{n}\rangle=&\sqrt{n}|\pm,\overrightarrow{n-1}\rangle\pm\alpha|\pm,\overrightarrow{n}\rangle,\\
    \hat{a}^\dagger|\pm,\overrightarrow{n}\rangle =& \sqrt{n+1}|\pm,\overrightarrow{n+1}\rangle\pm\alpha|\pm,\overrightarrow{n}\rangle,\\
    \hat{n}|\pm,\overrightarrow{n}\rangle =& \pm\alpha\sqrt{n+1}|\pm,\overrightarrow{n+1}\rangle + (n+\alpha^2)|\pm,\overrightarrow{n}\rangle\\&\pm\alpha\sqrt{n}|\pm,\overrightarrow{n-1}\rangle.
\end{aligned}
$$
Inverting the direction of arrows for the above gives the equation for $|\pm,\overleftarrow{n}\rangle$. Finally, for $\hat{R}$ and $\hat{R}^\dagger$, we have 
$$
\begin{aligned}
    \hat{R}|\pm,\overrightarrow{n}\rangle
    =& \mp\sqrt{n}|\mp,\overleftarrow{n-1}\rangle,\\
    \hat{R}|\pm,\overleftarrow{n}\rangle
    =& 2\alpha|\mp,\overrightarrow{n}\rangle\mp\sqrt{n}|\mp,\overrightarrow{n-1}\rangle,\\
    \hat{R}^\dagger|\pm,\overrightarrow{n}\rangle
    =& 2\alpha|\mp,\overleftarrow{n}\rangle\pm\sqrt{n+1}|\mp,\overleftarrow{n+1}\rangle,\\
    \hat{R}^\dagger|\pm,\overleftarrow{n}\rangle
    =& \pm\sqrt{n+1}|\mp,\overrightarrow{n+1}\rangle.
\end{aligned}
$$
Assuming subsequent collapse or measurement right after the jump, we describe the stochastic jump direction and probability between states in Fig. ~\ref{fig:Markov_diagram}. 

A pure state $|\psi\rangle$ can be a steady state of the dynamics generated by $\mathcal{D}[\hat{R}]$ if and only if $|\psi\rangle$ is a right eigenstate of $\hat{R}$ and, in addition, either the corresponding eigenvalue is zero or $|\psi\rangle$ is also the matching eigenstate of $\hat{R}^\dagger$. Now, observe that $\hat{R}^2 = \alpha^2-\hat{a}^2$ and $\hat{R}|\psi\rangle=\lambda|\psi\rangle \to \hat{R}^2|\psi\rangle=\lambda^2|\psi\rangle$. Since the only eigenspace satisfying the stationary condition of the two-photon jump operator is $\mathrm{span}(|\pm\alpha\rangle_b)$ with zero eigenvalue, our candidate of steady states reduces to elements of the  $\mathrm{span}(|\phi\rangle_s\otimes(c_+|\alpha\rangle_b+c_-|-\alpha\rangle_b))$. Calculating the action of $\hat{R}$ on such states reveals that the pure states can be steady only if they are in $\mathrm{span}(|\pm,\overrightarrow{0}\rangle)$.

\vspace{0.3cm}

\noindent \textbf{Table of experimental parameters.} We present the parameters used throughout the numerical analysis in Table~\ref{tab:params}, which are obtained from recent relevant experimental studies.
\begin{table}[ht]
    \centering
    \begin{ruledtabular}
        (a) Hyperfine qubits and motional modes of trapped-ions
        \begin{tabular}{p{5cm} r}
            Thermal noise ($\gamma_\mathrm{th}/2\pi$). & $0.014 - 4.7\ \mathrm{Hz}$ \\ 
            Bosonic dephasing ($\kappa_{\hat{n}}/2\pi$). &  $ 1.6 -\ 4.0\ \mathrm{Hz}$\\ 
            Qubit bit flip ($\kappa_{\hat{\sigma}_x}/2\pi$). & $\ <\ 0.03\ \mathrm{Hz}$ \\
            Qubit phase flip ($\kappa_{\hat{\sigma}_z}/2\pi$). & $<\ 0.27\ \mathrm{Hz}$ \\
            \hline
            Controlled beam-spliting ($g/2\pi$) & $0.31 - 0.87\ \mathrm{kHz}$\\
            Bath relaxation ($\gamma_b/2\pi$) & $\approx 13.4\ \mathrm{kHz}$\\
        \end{tabular}
        (b) Qubits and bosonic modes of the superconducting cavity
        \begin{tabular}{p{5cm} r}
             Photon loss ($\kappa'_\mathrm{th}/2\pi$) & $2.1 - 2.3\ \mathrm{kHz}$ \\
             Bosonic dephasing ($\kappa'_{\hat{n}}/2\pi$) & $0.37 - 0.73\ \mathrm{kHz} $ \\
             Qubit bit flip ($\kappa'_{\hat{\sigma}_x}/2\pi$) & $0.25-0.95\ \mathrm{kHz}$\\
             Qubit phase flip ($\kappa'_{\hat{\sigma}_z}/2\pi$) & $0.08-1.6\ \mathrm{kHz}$\\
             \hline
             3-wave mixing SNAIL ($g'/2\pi$) & $0.3-60 \ \mathrm{MHz}$\\
             Bath relaxation ($\gamma'_b/2\pi$) & $\approx10.7\ \mathrm{MHz}$\\
        \end{tabular}
    \end{ruledtabular}
    \caption{System parameters considered for the concatenated hybrid/cat qubit. We considered parameters related to the trapped-ion system~\cite{PhysRevLett.124.170502, jeon2025multimodebosonicstatetomography} (up). Relevant rates of the superconducting cavity and qubits (down) ~\cite{van_damme_advanced_2024, PhysRevX.15.011070, PhysRevApplied.11.054060, frattini_three-wave_2021}.}
    \label{tab:params}
\end{table}

\noindent \textbf{Displacement sensing with noise-suppressed hybrid states.}
For a given state $\hat{\rho}$, the QFI $\mathcal{F}_\mathrm{Q}(\hat{\rho})$ with respect to an observable $\hat{O}$ is defined~\cite{paris_quantum_2009}
$$
    \mathcal{F}_\mathrm{Q}(\hat{\rho})=2\sum_{k,l}\frac{(\lambda_k-\lambda_l)^2}{(\lambda_k+\lambda_l)}|\bra{k}\hat{O}\ket{l}|^2,
$$
where $\lambda_k$ and $\ket{k}$ are the eigenvalue and eigenvector of a given state $\hat{\rho}$. The summation goes for all $k$ and $l$ such that $\lambda_k+\lambda_l>0$. For a pure state $\ket{\psi}$, QFI reduces to the variance of the observable: $\mathcal{F}_\mathrm{Q}=4(\Delta\hat{O})^2=4(\langle\hat{O}^2\rangle_\psi-\langle\hat{O}\rangle_\psi^2)$. QFI designates the ultimate achievable precision in estimating the parameter $\beta$ using $\hat{\rho}_\beta:=e^{-i\beta\hat{O}}\hat{\rho}e^{i\beta\hat{O}}$ as a probe, which is the QCRB, as stated in the main text ($(\Delta\beta)^2\geq1/\mathcal{F}_\mathrm{Q}$). QCRB of certain quantum states can be lower than the standard quantum limit (SQL), the maximum precision achievable by a classical probe, indicating the possibility of parameter estimation with classically intractable precision.

For the displacement sensing, an entanglement between a discrete-variable (DV) system and a continuous-variable (CV) system can be utilized to obtain beyond-classical sensitivity~\cite{hempel_entanglement-enhanced_2013, gilmore_quantum-enhanced_2021}. For a pure entangled probe state $\ket{0}_\mathrm{L}:=\frac{1}{\sqrt{2}}(\ket{+}_s\otimes\ket{\alpha}_b+\ket{-}_s\otimes\ket{-\alpha}_b)$, the quantum Fisher information $\mathcal{F}_\mathrm{Q}$ with respect to the observable $\hat{X}:=\hat{a}+\hat{a}^\dagger$, the generator of $p$-displacement, can be readily obtained as $\mathcal{F}_\mathrm{Q}=4(\Delta\hat{X})^2=16\alpha^2+4$, which is larger than $\mathcal{F}_\mathrm{Q}(\ket{\alpha'}\bra{\alpha'})=4$, the maximum QFI for displacement sensing obtained from the coherent state probe.

\newpage

\maketitle

\onecolumngrid

\section{Supplemental Material for ``Autonomous Quantum Error Correction of Spin-Oscillator Hybrid Qubits''}

\section{I. Conserved quantities of $\mathcal{L}_\mathrm{R}$}

Observe that the hybrid parity, defined as following; $\hat{\Pi}_h:=\hat{\sigma}_z\otimes\hat{\Pi}_b$, where $\hat{\Pi}_b:=e^{i\pi\hat{n}}=\sum(-1)^n|n\rangle\langle n|_{b}$ is the bosonic parity operator, commutes with $\hat{R}$, $\hat{R}^\dagger$. Therefore, it is in kernel of the adjoint Liouvillian $\mathcal{L}_\mathrm{R}^\dagger$ ($\mathcal{L}_\mathrm{R}^\dagger\hat{\Pi}_h=0$) where the action of the adjoint Liouvillian $\cal L_\mathrm{R}^\dagger$ on an operator $\hat{J}$ can be formally written

\begin{equation}\label{eq:adj_Liouv}
    \mathcal{L}_\mathrm{R}^\dagger\hat{J}=\hat{R}^\dagger\hat{J}\hat{R}-\frac{1}{2}\{\hat{R}^\dagger\hat{R}, \hat{J}\}.
\end{equation}
As any state $\hat{\rho}$ satisfying $\mathcal{L}_\mathrm{R}\hat{\rho}=0$ is a steady state of the dynamics generated by $\mathcal{L}_\mathrm{R}$, any operator $\hat{J}$ satisfying $\mathcal{L}_\mathrm{R}^\dagger\hat{J}=0$ is a conserved quantity of the dynamics. Since the steady subspace of the dynamics is 4-dimensional (in the superoperator sense), an equal number of conserved quantities are expected~\cite{Sym_consv_VValbert}. Imposing the orthogonality condition given by the Hilbert-Schimidt inner product $\mathrm{Tr}[\hat{J}_{\mu\nu}^\dagger\hat{M}_{\mu'\nu'}]=\delta_{\mu\mu'}\delta_{\nu\nu'}$ to establish a correspondence between steady states and conserved quantities. The hybrid parity is connected to the two diagonal conserved quantities of the following:

$$
\begin{aligned}
    &\hat{J}_{00}:=\frac{1+\hat{\Pi}_h}{2}=\sum_{\mu\in\{0,1\}, n}|\mu\rangle\langle\mu|_{s}\otimes|2n+\mu\rangle\langle2n+\mu|_{b},\\
    &\hat{J}_{11}:=\frac{1-\hat{\Pi}_h}{2}=\sum_{\mu\in\{0,1\}, n}|\mu\rangle\langle\mu|_{s}\otimes|2n+1+\mu\rangle\langle2n+1+\mu|_{b}.
\end{aligned}
$$
The remaining conserved quantities cannot be generated from $\hat{\Pi}_h$. Let us define
$$
\hat R = \hat{\sigma}_y (\hat a - \alpha \hat \sigma_x) = i
\left(\begin{matrix}
    0 & -\hat b_+ \\
    \hat b_- & 0
\end{matrix}\right) = i
\left(\begin{matrix}
    \hat \Pi_b & 0 \\
    0 & \hat{\mathbb{I}}_b
\end{matrix}\right)
\left(\begin{matrix}
    0 & \hat b_- \\
    \hat b_- & 0
\end{matrix}\right)
\left(\begin{matrix}
    \hat{\mathbb{I}}_b & 0 \\
    0 & \hat \Pi_b
\end{matrix}\right),
$$
in a block matrix form in the x-basis $\hat{A}=\left(\begin{matrix}
    \hat{A}_{++} & \hat{A}_{+-} \\
    \hat{A}_{-+} & \hat{A}_{--}
\end{matrix}\right)$ where $\hat{A}_{ij}:=\mathrm{Tr}_s[\hat{A}(\ket{i}\bra{j}_s\otimes\hat{\mathbb I}_b)]$ are the bosonic operators obtained from projecting onto the spin-$x$ subspace and we define $\hat b_\pm = \hat a \pm \alpha = -\hat{\Pi}_b \hat b_\mp \hat{\Pi}_b$. The choice of $x$-basis is to better reflect the structure of $\hat{R}$. For the anticommutator part of Eq.~\eqref{eq:adj_Liouv},
$$
\hat R^\dagger \hat R =
\left(\begin{matrix}
    \hat{\mathbb{I}}_b & 0 \\
    0 & \hat{\Pi}_b
\end{matrix}\right)
\left(\begin{matrix}
    \hat b_-^\dagger \hat b_- & 0 \\
    0 & \hat b_-^\dagger \hat b_-
\end{matrix}\right)
\left(\begin{matrix}
    \hat{\mathbb{I}}_b & 0 \\
    0 & \hat{\Pi}_b
\end{matrix}\right)
= \left(\begin{matrix}
    \hat b_-^\dagger \hat b_- & 0 \\
    0 & \hat b_+^\dagger \hat b_+
\end{matrix}\right),
$$
which is a shifted number operator with their direction of shift conditioned on the spin-$x$ eigenvalue. 
$$
{\cal L}_R^\dagger\hat{J} =
\left(\begin{matrix}
    \hat{\mathbb{I}}_b & 0 \\
    0 & \hat{\Pi}_b
\end{matrix}\right)
\left(\begin{matrix}
    0 & \hat b_-^\dagger \\
    \hat b_-^\dagger & 0
\end{matrix}\right)
\left(\begin{matrix}
    \hat{\Pi}_b & 0 \\
    0 & \hat{\mathbb{I}}_b
\end{matrix}\right)
\hat{J}
\left(\begin{matrix}
    \hat{\Pi}_b & 0 \\
    0 & \hat{\mathbb{I}}_b
\end{matrix}\right)
\left(\begin{matrix}
    0 & \hat b_- \\
    \hat b_- & 0
\end{matrix}\right)
\left(\begin{matrix}
    \hat{\mathbb{I}}_b & 0 \\
    0 & \hat{\Pi}_b
\end{matrix}\right)
-\frac{1}{2} \hat{J} \left(\begin{matrix}
    \hat b_-^\dagger \hat b_- & 0 \\
    0 & \hat b_+^\dagger \hat b_+
\end{matrix}\right) - \frac{1}{2}\left(\begin{matrix}
    \hat b_-^\dagger \hat b_- & 0 \\
    0 & \hat b_+^\dagger \hat b_+
\end{matrix}\right) \hat{J} = 0.
$$
By conjugation of controlled-parity unitary $\left(\begin{matrix}
    \hat{\mathbb{I}}_b & 0 \\
    0 & \hat{\Pi}_b
\end{matrix}\right)$, we define $\hat{\tilde J}  
= \left(\begin{matrix}
    \mathbb{I}_b & 0 \\
    0 & \hat{\Pi}_b
\end{matrix}\right)
\hat{J}
\left(\begin{matrix}
    \mathbb{I}_b & 0 \\
    0 & \hat{\Pi}_b
\end{matrix}\right)
=
\left(\begin{matrix}
    \hat{\Pi}_b & 0 \\
    0 & \hat{\Pi}_b
\end{matrix}\right)
\left(\begin{matrix}
    \hat{\Pi}_b & 0 \\
    0 & \mathbb{I}_b
\end{matrix}\right)
\hat{J}
\left(\begin{matrix}
    \hat{\Pi}_b & 0 \\
    0 & \hat{\mathbb{I}}_b
\end{matrix}\right)
\left(\begin{matrix}
    \hat{\Pi}_b & 0 \\
    0 & \hat{\Pi}_b
\end{matrix}\right),
$
we have 
$$
\left(\begin{matrix}
    0 & \hat b_-^\dagger \\
    \hat b_-^\dagger & 0
\end{matrix}\right)
\left(\begin{matrix}
    \hat{\Pi}_b & 0 \\
    0 & \hat{\Pi}_b
\end{matrix}\right)
\hat{\tilde J}
\left(\begin{matrix}
    \hat{\Pi}_b & 0 \\
    0 & \hat{\Pi}_b
\end{matrix}\right)
\left(\begin{matrix}
    0 & \hat b_- \\
    \hat b_- & 0
\end{matrix}\right)
-\frac{1}{2} \hat{\tilde J} \left(\begin{matrix}
    \hat b_-^\dagger \hat b_- & 0 \\
    0 & \hat b_-^\dagger \hat b_-
\end{matrix}\right) - \frac{1}{2}\left(\begin{matrix}
    \hat b_-^\dagger \hat b_- & 0 \\
    0 & \hat b_-^\dagger \hat b_-
\end{matrix}\right) \hat{\tilde J} = 0.
$$
We then apply the Hadamard transformation $\hat{H} = \frac{1}{\sqrt{2}} \left(\begin{matrix}
    \hat{\mathbb{I}}_b & \hat{\mathbb{I}}_b \\
    \hat{\mathbb{I}}_b & -\hat{\mathbb{I}}_b\\
\end{matrix}\right)$, which leads to
$$
\left(\begin{matrix}
    \hat b_-^\dagger & 0 \\
    0 & - \hat b_-^\dagger
\end{matrix}\right)
\left(\begin{matrix}
    \hat{\Pi}_b & 0 \\
    0 & \hat{\Pi}_b
\end{matrix}\right)
(\hat{H} \hat{\tilde J} \hat{H})
\left(\begin{matrix}
    \hat{\Pi}_b & 0 \\
    0 & \hat{\Pi}_b
\end{matrix}\right)
\left(\begin{matrix}
    \hat b_- & 0 \\
    0 & - \hat b_-
\end{matrix}\right)
-\frac{1}{2} (\hat{H} \hat{\tilde J} \hat{H}) \left(\begin{matrix}
    \hat b_-^\dagger \hat b_- & 0 \\
    0 & \hat b_-^\dagger \hat b_-
\end{matrix}\right) - \frac{1}{2}\left(\begin{matrix}
    \hat b_-^\dagger \hat b_- & 0 \\
    0 & \hat b_-^\dagger \hat b_-
\end{matrix}\right) (\hat{H} \hat{\tilde J} \hat{H}) = 0.
$$
By defining
$H \tilde J H =
\left(\begin{matrix}
    \hat{A} & \hat{B} \\
    \hat{C} & \hat{D}
\end{matrix}\right)
$, we have the following four independent equations:
$$
\begin{aligned}
\hat b_-^\dagger \hat{\Pi}_b \hat{A} \hat{\Pi}_b \hat b_- - \frac{1}{2} \{ \hat b_-^\dagger \hat b_-, \hat{A}\}  &= 0 \\
\hat b_-^\dagger \hat{\Pi}_b \hat{B} \hat{\Pi}_b \hat b_- + \frac{1}{2} \{ \hat b_-^\dagger \hat b_-, \hat{B}\}  &= 0 \\
\hat b_-^\dagger \hat{\Pi}_b \hat{C} \hat{\Pi}_b \hat b_- + \frac{1}{2} \{ \hat b_-^\dagger \hat b_-, \hat{C}\}  &= 0 \\
\hat b_-^\dagger \hat{\Pi}_b \hat{D} \hat{\Pi}_b \hat b_- - \frac{1}{2} \{ \hat b_-^\dagger \hat b_-, \hat{D}\}  &= 0.
\end{aligned}
$$
Solving for diagonal equations, we have $\hat{A} \propto \mathbb{\hat I}_b$ and  $\hat{D} \propto \mathbb{\hat I}_b$ since the only conserved quantity of photon loss channel is $\hat{\mathbb{I}}_\mathrm{b}$, the trivial solution, corresponding to the trivial unique steady state $\ket{0}\bra{0}_b$ of the photon loss channel. We then take the following two linear combinations: $\hat{H} \hat{\tilde J} \hat{H} = \left(\begin{matrix}
    \mathbb{\hat I}_b & 0 \\
    0 & \mathbb{\hat I}_b
\end{matrix}\right) = \mathbb{\hat I}_s \otimes \mathbb{\hat I}_b$ and $\left(\begin{matrix}
    \mathbb{\hat I}_b & 0 \\
    0 & -\mathbb{\hat I}_b
\end{matrix}\right) = \hat \sigma_x \otimes \mathbb{\hat I}_b$
to obtain the conserved quantities $\hat{J}_0 = \left(\begin{matrix}
    \mathbb{\hat I}_b & 0 \\
    0 & \mathbb{\hat I}_b
\end{matrix}\right)$ and $\hat{J}_3 = \left(\begin{matrix}
    0 & \hat{\Pi}_b \\
    \hat{\Pi}_b & 0
\end{matrix}\right) = \hat \sigma_z \otimes \hat{\Pi}_b$ which give aforementioned diagonal conserved quantities $\hat{J}_{00}$, $\hat{J}_{11}$. The remaining two solutions are in the form of 
$\hat{J}_1 = \left(\begin{matrix}
    \hat{K} & 0 \\
    0 & -\hat{\Pi}_b \hat{K} \hat{\Pi}_b
\end{matrix}\right)$ and $\hat{J}_2=\left(\begin{matrix}
    0 & -\hat{K}\hat{\Pi}_b \\
    \hat{\Pi}_b \hat{K} & 0
\end{matrix}\right)=\hat{J}_3\hat{J}_1=-\hat{J}_1\hat{J}_3$ where $\hat{K}$ is the solution of the off-diagonal equation, taken to be Hermitian without loss of generality, given by 
\begin{equation}\label{eqnK}
    \hat b_-^\dagger \hat{\Pi}_b \hat{K} \hat{\Pi}_b \hat b_- + \frac{1}{2} \{ \hat b_-^\dagger \hat b_-, \hat{K}\}  = 0.
\end{equation}
Taking $\hat B$, $\hat C$ to be $\hat{K}$ and taking their sum (subtraction) gives $\hat J_1$ ($\hat{J}_2$). Orthogonality relations regarding off-diagonal quantities write
$$
    \mathrm{Tr}[\hat{J}_1^\dagger\ket{0}\bra{1}_\mathrm{L}]=\mathrm{Tr}[\hat{J}_1^\dagger\ket{1}\bra{0}_\mathrm{L}]=  \mathrm{Tr}[\hat{J}_2^\dagger\ket{0}\bra{1}_\mathrm{L}]=
    -\mathrm{Tr}[\hat{J}_2^\dagger\ket{1}\bra{0}_\mathrm{L}]=\bra{\alpha}\hat{K}\ket{\alpha}_b,
$$
fixing $\bra{\alpha}\hat{K}\ket{\alpha}_b=1$ without loss of generality, we obtain the remaining off-diagonal conserved quantities $\hat{J}_{01}=\frac{1}{2}(\hat{J}_1+\hat{J}_2)$ and $\hat{J}_{10}=\frac{1}{2}(\hat{J}_1-\hat{J}_2)$.

\section{II. Logical error rates of physical noise}

The error process $\mathcal{L}_\mathrm{E}$ perturbatively manifests and induces the decay of the population of the initial state prepared to be the logical eigenstate. The symmetric transition rate $\gamma_i$ between $|\psi^i_0\rangle$ and $|\psi^i_1\rangle$, the eigenstates of $\hat{\sigma}_i$, is given by the rate equation below:
$$
\begin{aligned}
    \dot p^i_0 &= - \gamma_i p^i_0 + \gamma_i p^i_1 ,\\
    \dot p^i_1 &= \gamma_i p^i_0 - \gamma_i p^i_1.
\end{aligned}
$$
Here, $p^i_j:=\langle\psi^i_j|\hat{\rho}|\psi^i_j\rangle$ are population of each eigenstate for a given ensemble $\hat{\rho}$. Explicitly solving for $t$ yields $p_j^i(t) = 1/2+(p_j^i(0)-1/2)e^{-2\gamma_i t}$.

Conserved quantities of the recovery dynamics can be used to perturbatively obtain the logical error rate for a given physical error channel $\kappa_\mathrm{E}\mathcal{D}[\hat{E}]$, as they produce the population of the corresponding logical eigenstate in a given $\hat \rho(t)$, by taking the inner product:
$$
\begin{aligned}
    \gamma_\mathrm{Z}(\hat{E})\approx&-\bigg(\frac{d}{dt}\mathrm{Tr}[\hat{J}_{++}^\dagger e^{(\mathcal{L}_\mathrm{R}+\mathcal{L}_\mathrm{E})t}\ket{+}\bra{+}_\mathrm{L}]\bigg)\bigg|_{t=0}=-\kappa_\mathrm{E}\mathrm{Tr}[\hat{J}_{++}^\dagger\mathcal{D}[\hat{E}](\ket{+}\bra{+}_\mathrm{L})],\\
    \gamma_\mathrm{X}(\hat{E})\approx&-\bigg(\frac{d}{dt}\mathrm{Tr}[\hat{J}_{00}^\dagger e^{(\mathcal{L}_\mathrm{R}+\mathcal{L}_\mathrm{E})t}\ket{0}\bra{0}_\mathrm{L}]\bigg)\bigg|_{t=0}=-\kappa_\mathrm{E}\mathrm{Tr}[\hat{J}_{00}^\dagger\mathcal{D}[\hat{E}](\ket{0}\bra{0}_\mathrm{L})],
\end{aligned}
$$
respectively, and $\hat{J}_{++}:=\frac{1}{2}(\hat{J}_0+\hat{J}_1)$ satisfying $\mathrm{Tr}[\hat{J}_{++}^\dagger\ket{+}\bra{+}_\mathrm{L}]=1$. We start with the spin-z noise, the simplest error channel:
$$
\gamma_Z(\hat{\sigma}_z)=\frac{\kappa_{\hat{\sigma}_z}}{2}(1+\bra{-\alpha}\hat{K}\ket{-\alpha}_b)
$$
Then, evaluating $\bra{-\alpha} K \ket{-\alpha}$ is enough for calculating the first-order approximation of the error rate on the code space. We start from the Eq.\eqref{eqnK} in the displaced frame ($\hat{A}'=\hat{D}^\dagger(\alpha)\hat{A}\hat{D}(\alpha)$).
$$
\hat{a}^\dagger\hat{\Pi}_b'\hat{K}'\hat{\Pi}_b'\hat{a}+\frac{1}{2}\{\hat{a}^\dagger\hat{a},\hat{K}'\}=0
$$
Note that boundary terms other than $\bra{0}\hat{K}'|0\rangle=\bra{\alpha}\hat{K}\ket{\alpha}=1$ banish $\bra{m>0}\hat{K}'\ket{0}=\bra{0}\hat{K}'\ket{n>0}=0$. Let us define $F(z,w):=\bra{z}\hat{K}'\ket{w}_B=e^{\frac{1}{2}(|z|^2+|w|^2)}\bra{z}\hat{K}'\ket{w}$, where $\ket{z}_B:=e^{\frac{|z|^2}{2}}|z\rangle_b$ is the unnormalized coherent state~\cite{bargmann_hilbert_1961} satisfying $\hat{a}\ket{ z}_B=z\ket{z}_B$, $\hat{a}^\dagger\ket{z}_B=\partial_z\ket{z}_B$. The corresponding differential relation regarding the above equation is now written
$$
zwe^{-4\alpha^2-2\alpha(z+w)}F(-2\alpha-z,-2\alpha-w)+\frac{1}{2}(z\partial_z+w\partial_w)F(z,w)=0.
$$
Introducing the relevant single variable function $h(x):=e^{-x^2}F(x,x)=\bra{x}\hat{K}'\ket{x}_b$,
$$
\frac{d}{dx}h(x)+2xh(x)=-2xh(-2\alpha-x),
$$
since $\frac{d}{dx}F(x,x)=(\partial_z+\partial_w)F(z,w)\big|_{z=w=x}$. Defining $S(x):=h(x)+h(-2\alpha-x)$,
$$
\frac{d}{dx}S(x)=-4(x+\alpha)S(x)\Rightarrow S(x)=e^{-4\alpha x-2x^2}S(0).
$$
Where $S(0)=h(0)+h(-2\alpha)=1+\bra{-\alpha}\hat{K}\ket{-\alpha}$. Now evaluating $\int_0^{-2\alpha}dxh'(x)$ gives
$$
\bra{-\alpha}\hat{K}\ket{-\alpha}_b-1=-2(1+\bra{-\alpha}\hat{K}\ket{-\alpha}_b)\int_0^{-2\alpha}dx (xe^{-4\alpha x-2x^2}).
$$
Solving for $\bra{-\alpha}\hat{K}\ket{-\alpha}$, we arrive at
\begin{equation}\label{eqn:neg_expct}
\bra{-\alpha}\hat{K}\ket{-\alpha}_b=\frac{1-2I(\alpha)}{1+2I(\alpha)},
\end{equation}
where 
$$
\begin{aligned}
    I(\alpha):=&\int_0^{-2\alpha}dx(xe^{-4\alpha x-2x^2})=\alpha\sqrt{\frac{\pi}{2}}e^{2\alpha^2}\mathrm{erf}(\sqrt{2}\alpha), \\\mathrm{erf}(x):=&\frac{2}{\sqrt{{\pi}}}\int_{0}^{x}dte^{-t^2},
\end{aligned}
$$
concluding derivation of $\gamma_\mathrm{Z}(\hat{\sigma}_z)=\kappa_{\hat{\sigma}_z}/(1+2I(\alpha))$. Note that $\gamma_\mathrm{X}(\hat{\sigma}_z)=0$, since $\hat{J}_{00}^\dagger\mathcal{D}[\hat{\sigma}_z]\hat{\rho}=\mathcal{D}[\hat{\sigma}_z](\hat{J}_{00}^\dagger\hat{\rho})$ and trace of dissipator is always zero. 

For spin-x noise, $\mathcal{D}[\hat{\sigma}_x](\ket{0}\bra{0}_\mathrm{L})=\ket{1}\bra{1}_\mathrm{L}-\ket{0}\bra{0}_\mathrm{L}$ yields $\gamma_\mathrm{X}(\hat{\sigma}_x)=\kappa_{\hat{\sigma}_x}$, and $\gamma_\mathrm{Z}(\hat{\sigma}_x)=0$ since $\mathcal{D}[\hat{\sigma}_x](\ket{+}\bra{+}_\mathrm{L})=0$.

\begin{figure}[t]
    \centering
    \includegraphics[width=0.6\linewidth]{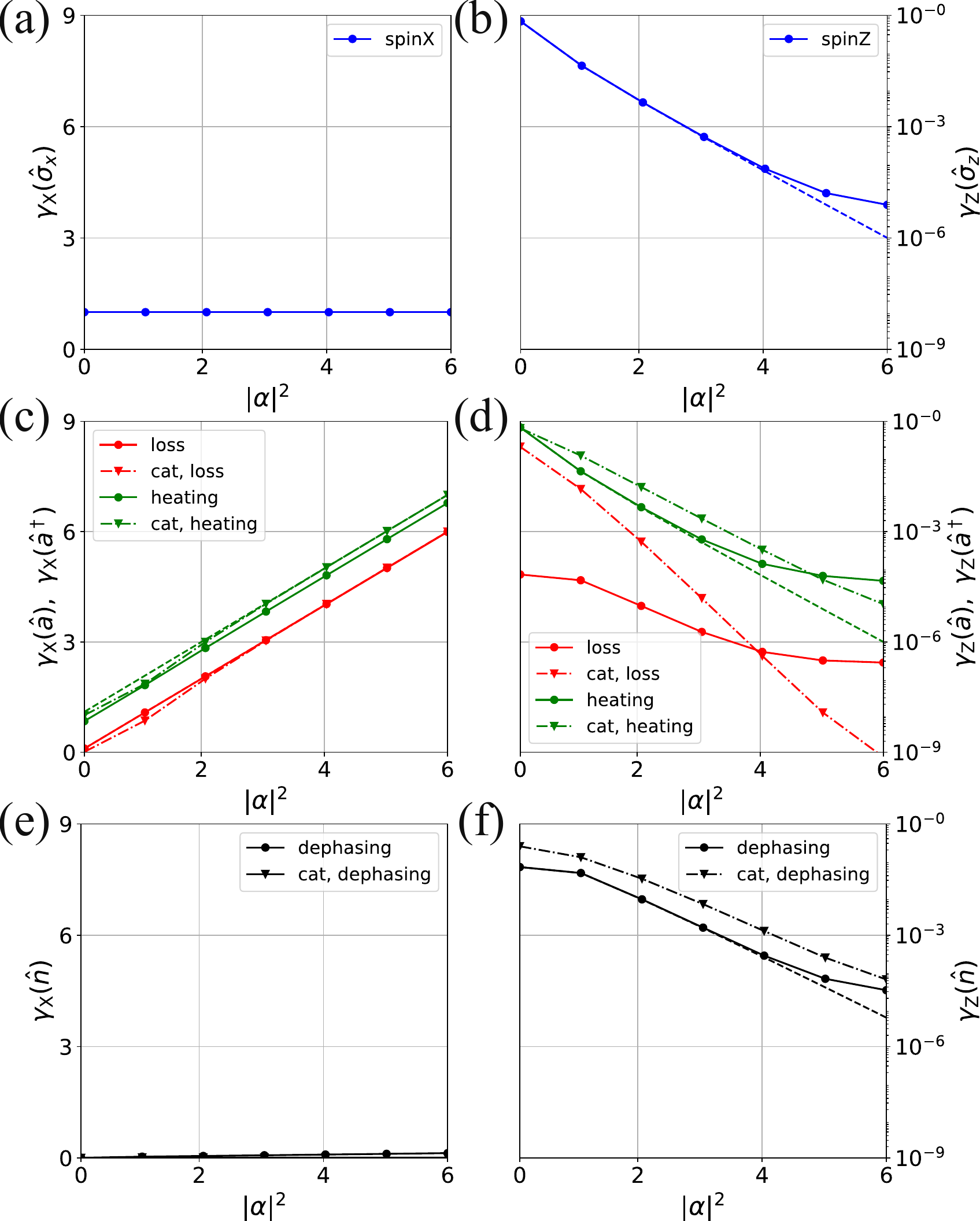}
    \caption{Logical error rates with respect to the state mean photon number $|\alpha|^2$. Simulations (real lines) are obtained in a rapid dissipation limit ($\kappa_\mathrm{R}/\kappa_\mathrm{E}=10^4$). Theoretical estimates (dashed) derived in Section II and simulation results of dynamically stabilized cat states under equivalent parameters (dot-dashed) are also presented for comparison.}
    \label{fig:errors}
\end{figure}

For loss, 
$$
\begin{aligned}
    \gamma_\mathrm{X}(\hat{a})=&-\kappa_{\hat{a}}\mathrm{Tr}[\hat{J}_{00}^\dagger\mathcal{D}[\hat{a}](\ket{0}\bra{0}_\mathrm{L})]\\
    =&-\frac{\kappa_{\hat{a}}}{4}\mathrm{Tr}[\mathcal{D}[\hat{a}](\ket{\alpha}\bra{\alpha}_b+\ket{-\alpha}\bra{-\alpha}_b)+\hat{\Pi}_b\mathcal{D}[\hat{a}](\ket{\alpha}\bra{-\alpha}_b+\ket{-\alpha}\bra{\alpha}_b)]\\=&\kappa_{\hat a}\alpha^2,\\
    \gamma_\mathrm{Z}(\hat{a})=&-\kappa_{\hat{a}}\mathrm{Tr}[\hat{J}_{++}^\dagger\mathcal{D}[\hat{a}](\ket{+}\bra{+}_\mathrm{L})]\\
    =&\frac{1}{2}\bigg(\alpha^2-\frac{1}{2}\mathrm{Tr}[\hat{K}\{\hat{a}^\dagger\hat{a},\ket{\alpha}\bra{\alpha}_b\}]\bigg)\\
    =&0,
\end{aligned}
$$
where we used $\bra{\alpha}\{\hat{K},\hat{a}^\dagger\hat{a}\}\ket{\alpha}=2\alpha^2$, obtained from Eq.~\eqref{eqnK} and the boundary conditon on $\hat{K}$, $\bra{m>0}\hat{K}'\ket{0}=\bra{0}\hat{K}'\ket{n>0}=0$.
$$
\bra{\alpha}\hat{K}\hat{a}^\dagger\hat{a}\ket{\alpha}_b=\langle0|\hat{K}'(\hat{a}^\dagger+\alpha)(\hat{a}+\alpha)\ket{0}_b=\alpha^2,
$$and the same for $\hat{a}^\dagger\hat{a}\hat{K}$. Logical error rates from heating can be similarly obtained
$$
\begin{aligned}
    \gamma_\mathrm{X}(\hat{a}^\dagger)=&-\kappa_{\hat{a}^\dagger}\mathrm{Tr}[\hat{J}_{00}^\dagger\mathcal{D}[\hat{a}^\dagger](\ket{0}\bra{0}_\mathrm{L})]=\kappa_{\hat{a}^\dagger}(\alpha^2+1),\\
    \gamma_\mathrm{Z}(\hat{a}^\dagger)=&-\kappa_{\hat{a}^\dagger}\mathrm{Tr}[\hat{J}_{++}^\dagger\mathcal{D}[\hat{a}^\dagger](\ket{+}\bra{+}_\mathrm{L})]=\frac{1}{1+2I(\alpha)},
\end{aligned}
$$
where we used Eq.\eqref{eqn:neg_expct}. For bosonic dephasing,
$$
\begin{aligned}
    \gamma_\mathrm{X}(\hat n) =&-\kappa_{\hat{n}}\mathrm{Tr}[\hat{J}_{00}^\dagger\mathcal{D}[\hat n]\ket{0}\bra{0}_\mathrm{L}]\\
    =&-\kappa_{\hat{n}}\mathrm{Tr}[\mathcal{D}[\hat{n}](\hat{J}_{00}^\dagger\ket{0}\bra{0}_\mathrm{L})]\\
    =&0,\\
    \gamma_\mathrm{Z}(\hat n) =&-\kappa_{\hat{n}}\mathrm{Tr}[\hat{J}_{++}^\dagger\mathcal{D}[\hat{n}]\ket{+}\bra{+}_\mathrm{L}]\\
    =&-\kappa_{\hat n}\mathrm{Tr}[\hat K\mathcal{D}[\hat{n}](\ket{\alpha}\bra{\alpha}_b)]\\
    =&\frac{\kappa_{\hat n}\alpha^2}{1+2I(\alpha)}.
\end{aligned}
$$
For the last equality in the derivation of $\gamma_\mathrm{Z}(\hat n)$, we used $\bra{\alpha}\hat{n}\hat{K}\hat{n}\ket{\alpha}_b=\alpha^4+\alpha^2\bra{0}\hat{a}\hat{K}'\hat{a}^\dagger\ket{0}_b=\alpha^4-\alpha^2\langle-\alpha|\hat{K}|-\alpha\rangle_b=\alpha^4-\alpha^2\frac{1-2I(\alpha)}{1+2I(\alpha)}$ and $\bra{\alpha}\hat{K}\hat{n}^2\ket{\alpha}_b=\bra{\alpha}\hat{n}^2\hat{K}\ket{\alpha}_b=\alpha^4+\alpha^2$, where $\bra{0}\hat{a}\hat{K}'\hat{a}^\dagger\ket{0}_b=-\langle-\alpha|\hat{K}|-\alpha\rangle_b$ can be obtained recursively from Eq.~\eqref{eqnK}.
Figure ~\ref{fig:errors} shows the simulation result of AutoQEC dynamics in the rapid dissipation limit ($\kappa_\mathrm{R}/\kappa_\mathrm{E}=10^4$), which agrees well with the perturbatively derived results.

\end{document}